\documentclass[12pt]{article}

\usepackage{cyclones}

\usepackage[backend=biber,style=numeric-comp,sorting=none,autopunct=true,language=USenglish]{biblatex}
\providecommand{\nada}[1]{}

\newcommand{\C}[1]{\ensuremath{{\text{C}#1}}}
\newcommand{\N}[1]{\ensuremath{\text{N}_{\text{\C{#1}}}}}

\newcommand{\nk}[1]{\N{#1}/\N{1+}}
\newcommand{\dn}[1]{\Delta\!\left(\frac{\N{#1}}{\N{1+}}\right)}

\renewcommand{\section}[1]{\needspace{5\baselineskip}\bigskip\noindent\textbf{\uline{#1}}\medskip\par}
\renewcommand{\subsection}[1]{\needspace{4\baselineskip}\medskip\noindent\textbf{#1}\smallskip\par}

\begin{document}

\title{Genuine Increases in Tropical Cyclone Intensities}

\author{Ivo Welch\\UCLA\thanks{Email: \texttt{ivo.welch@ucla.edu}. ORCID: \texttt{0000-0002-4347-7250}. The ADT-HURSAT dataset was created by the NOAA/NCEI satellite team, principally the team of James~P.\ Kossin and Kenneth~R.\ Knapp.}}

\date{March 2026}

\maketitle

\begin{abstract}
  \textcite{kossin2020global} report a rising ratio of satellite observations of major \C3--\C5 storms relative to all \C1--\C5 storms from 1979 to 2017.  Decomposing their $R = \N{3+}/\N{1+}$ statistic into per-category shares shows that their trend was driven primarily by fewer \C1 rather than more \C3--\C5 observations.  From the first half to the second half of their sample period, their per-year \C1 observations fell by 17\%.  However, extending the record through 2023 greatly changes the picture.  Although the relative decline in \C1 observations persists, \C3 and \C4 observations now increase, too.  The signal about the intensification of storms now becomes genuine in the extended sample, in that it is driven no longer only by fewer weak but now also by more strong tropical cyclone observations.
\end{abstract}

\clearpage

\section{Introduction}

A broad literature has documented intensification of tropical cyclones \parencite{emanuel2005increasing,webster2005changes,elsner2008increasing,holland2014recent,knutson2020tropical}. Most recently, \textcite{kossin2020global} construct and analyze a satellite-based tropical cyclone intensity dataset designed for temporal homogeneity (the ADT-HURSAT v6 1979--2017 record).  Specifically, their analyses rest on their ratio statistic, here renamed from ``$P_\text{Maj}$'' to ``R'':
\begin{eq*}
  R &\equiv& \frac{\N{3+}}{\N{1+}} &\equiv& \frac{\text{\C3+\C4+\C5 observations}}{\text{\C1+\C2+\C3+\C4+\C5 observations}} \;,
\end{eq*}
where ``\C{k}'' denotes Saffir-Simpson category~$k$,\footnote{The standard thresholds are \C1 ($64$--$82$~kt), \C2 ($83$--$95$~kt), \C3 ($96$--$112$~kt), \C4 ($113$--$136$~kt), \C5 ($\ge 137$~kt).} and each $N$ counts individual satellite observations classified by wind speed at the time of the satellite overpass.\footnote{This brief report will henceforth often call storm observations simply storms.  However, a single cyclone can contribute many observations as it strengthens and weakens.  The two measures are different because the per-satellite observation record of $R$ also reflects faster promotion of cyclones in recent years (from a median of 18h at C1 in the late 20th century to 12h in the early 21st century). \nada{When the analysis is conducted with (the maximum strength of) individual storms rather than individual satellite measurements, the $\Delta R$ statistic becomes statistically insignificant.  ($\Delta \N4/\N{1+}$ still increases and remains statistically significant. However, $\Delta \N3/\N{1+}$ no longer increases, at all.)}}  The $R$ statistic measures the incidence of \C3, \C4, and \C5 cyclone observations relative to \C1\ and \C2 observations.  Their interest is in comparing the change of $R$ over time.  The paper reports (1)~a higher $R$ in the second half of their sample and (2) a positive global trend of $R$ at approximately $+6$\% per decade (Mann-Kendall $P = 0.02$).

\textcite{kossin2020global} interpret their rising ratio as ``an increase in environmental potential intensity [that] is expected to manifest as a shift in the TC intensity distribution toward greater intensity and an increase in mean intensity.''   A natural reading of such an interpretation is that the rising ratio reflects weaker cyclones being pushed into stronger versions of themselves.  However, this need not necessarily be so, because there are not only numerator but also denominator effects at work.\footnote{They state that they focus on the ratio because it reduces dependence on absolute measures, possibly improving robustness.  While true, it does not negate the denominator effect --- unless one views the reduction in weak storms as just as important as the increase in strong storm in terms of an effect --- or why non-detection of weak storms would be more pronounced later in the sample.}

The paper here uses newer NCEI ADT-HURSAT~v7 data to update the perspective and disentangle the role of increases in strong cyclones from that of decreases in weak cyclones.  Writing the Kossin $R$ ratio as $R = \nk{3} + \nk{4} + \nk{5}$, where \N{1+} is the total number of cyclone-strength observations, its temporal change can be decomposed exactly as
\begin{eq*}
  \Delta R &\equiv& \dn{3} + \dn{4} + \dn{5} &\equiv& -\,\dn{1} - \dn{2} \;.
\end{eq*}
Using these identities shows that the increase in $R$ in \textcite{kossin2020global} was a phenomenon primarily of decreasing \C1 rather than increasing \C{3+} incidence.  However, the paper here documents that in the extended sample through 2023,\footnote{The tabulated results cover 1979--2023 (the last complete 3-year period).  Extending the annual (non-triad) analysis through 2024 does not change any conclusion.} this is no longer the case.  Instead, there are now clear observational increases in the incidences of high-intensity cyclones, specifically \C3\ and \C4\ cyclones, too.

\bigskip

\section{Results}\label{sec:results}

\subsection{Early vs.\ Late Comparison}

This paper uses the NCEI's updated and publicly posted ADT-HURSAT~v7b dataset. Their data were constructed using an improved Advanced Dvorak Technique (ADT) version~9.0 on HURSAT~v07b satellite imagery.  It covers 4,852 storms across six basins (NA, EP, WP, NI, SI, SP) from 1978--2025, with approximately 265,000 observations at approximately 3-hourly resolution and 1-kt wind speed values.  The Materials and Methods appendix summarizes the updated data set and how it differs from the earlier version used in \textcite{kossin2020global}.  Both data sets were constructed by the NCEI team.

\instbl{tbl:counts}

\begin{table}[ht]
  \small

\vspace{-2em}

  \tblcaption{tbl:counts}{Early vs Late Cyclone HURSAT v7 Observations}

  \panel{A}{Observation Counts}

  \begin{ctabular}{l rrssR ssss rrssR}
    \toprule
           & \mc3c{Kossin Years} & \mc3c{Extended Years} \\
    \cmidrule(lr){2-4} \cmidrule(lr){5-7}
    Cat    & \mc1c{1979-1997} & \mc1c{1998-2017} & \mc1{csss}{$\Delta N$} & \mc1c{1979-2000} & \mc1c{2001-2023} & \mc1c{$\Delta N$} \\
    \midrule
    \C1    & 6,477  & 5,638  & -839 & 7,301  & 6,489  & -812 \\
    \C2    & 2,625  & 2,679  & +54  & 3,001  & 3,294  & +293 \\
    \addlinespace
    \C3    & 3,752  & 3,860  & +108 & 4,218  & 4,786  & +568 \\
    \C4    & 2,521  & 2,636  & +115 & 2,795  & 3,417  & +622 \\
    \C5    &   621  &   492  & -129 &   664  &   594  & -70  \\
    \midrule
    \C{1+} & 15,996 & 15,305 & -691 & 17,979 & 18,580 & +601 \\
    \bottomrule
  \end{ctabular}

  \panel{B}{Per-Year Rates}

  \begin{ctabular}{l rr RR ssss rr RR}
    \toprule
           & \mc4{cs}{Kossin Years}  & \mc4c{Extended Years} \\
    \cmidrule(lr){2-5} \cmidrule(lr){6-9}
    Cat    & \mc1c{79-97} & \mc1c{98-17} & \mc1{r}{$\Delta$/yr} & \mc1{css}{$\%\Delta$/yr}
    & \mc1c{79-00} & \mc1c{01-23} & \mc1r{$\Delta$/yr}  & \mc1{c}{$\%\Delta$/yr}\\
    \midrule
    \C1    & 341 & 282 & -59  & -17\% & 332 & 282 & -50 & -15\% \\
    \C2    & 138 & 134 & -4   & -3\%  & 136 & 143 & +7  & +5\%  \\
    \addlinespace
    \C3    & 197 & 193 & -5   & -2\%  & 192 & 208 & +16 & +9\%  \\
    \C4    & 133 & 132 & -1   & -1\%  & 127 & 149 & +22 & +17\% \\
    \C5    &  33 &  25 & -8   & -25\% &  30 &  26 & -4  & -14\% \\
    \midrule
    \C{1+} & 842 & 765 & -77  & -9\%  & 817 & 808 & -9  & -1\%  \\
    \bottomrule
  \end{ctabular}

  \panel{C}{Share of Categories in all \C{1+} Categories ($\N{k}/\N{1+}$), Comprising $R$ Statistic}

  \begin{ctabular}{c rrR ssrrR}
    \toprule
    & \multicolumn{3}{c}{Kossin Years} & \multicolumn{3}{c}{Extended Years}                \\
    \cmidrule(lr){2-4} \cmidrule(lr){5-7}
    Cat & 79-97 & 98-17 & $\Delta$ & 79-00 & 01-23 & $\Delta$ \\
    \midrule
    \C1 & 0.405 & 0.368 & -0.037 & 0.406 & 0.349 & -0.057 \\
    \C2 & 0.164 & 0.175 & +0.011 & 0.167 & 0.177 & +0.010 \\
    \addlinespace
    \C3 & 0.235 & 0.252 & +0.018 & 0.235 & 0.258 & +0.023 \\
    \C4 & 0.158 & 0.172 & +0.015 & 0.156 & 0.184 & +0.028 \\
    \C5 & 0.039 & 0.032 & -0.007 & 0.037 & 0.032 & -0.005 \\
    \midrule
    \addlinespace
    \C3+\C4+\C5 &\multicolumn{2}{c}{Kossin $\Delta R^{\ddag}$} & +0.026 & & & +0.046 \\
    \bottomrule
  \end{ctabular}
  {\vspace{-1em}\scriptsize\hspace{7em}
    $^\ddag$~$\Delta R \equiv \Delta(\N3/\N{1+}) + \Delta(\N4/\N{1+}) + \Delta(\N5/\N{1+}) \equiv -\Delta(\N1/\N{1+}) - \Delta(\N2/\N{1+})$.}

  \smallskip

  \explain{Each period is split near its midpoint (1979--2017 at 1997/98, following Kossin; 1979--2023 at 2000/01). Panel~A shows raw totals.  Panel~B converts to per-year rates (count/year in each half: 19 vs.\ 20 for the Kossin period, 22 vs.\ 23 for the extended period).  Panel~C shows each category's share of all \C{1+} observations ($\N{k}/\N{1+}$).  Kossin's $P_\text{maj}$ is $R = \N{3+}/\N{1+}$.  $\Delta R$ sums the \C3--\C5 share changes.}
\end{table}

Table~\ref{tbl:counts} shows satellite observation counts by Saffir-Simpson category for the early and late halves of both the Kossin period (1979--2017, split at 1997/98) and the extended record (1979--2023, split at 2000/01).  Because the half-periods differ in length (19 vs.\ 20 and 22 vs.\ 23~years), Panel~B reports per-year rates alongside the raw counts in Panel~A.

As noted, $R$ can increase not only when there are more major cyclones but also when there are fewer weak cyclones.  The intensification interpretation would be more consistent with a stable or increasing total number of cyclone-strength observations.\footnote{There is also a larger reservoir of weaker tropical storms ready to be pushed into the \C1 category, too.}  Instead, in the Kossin sample, per-year \C{1+} observations fell from 842 to 765, a decline of 77 per year ($-9$\%).\footnote{The decline is even steeper in Kossin's original v6 data: \C1 per-year rates fell by 21\% (from 245 to 192/yr) and total \C{1+} by 14\% (from 504 to 434/yr), compared with $-17$\% and $-9$\% in v7.  The larger v6 decline reflects 6-hourly subsampling and 5-kt binning, which makes the \C1 bucket more sensitive to boundary shifts.}  The \C1 category alone accounts for 59 of those 77 (77\% of the total decline); every other category also declined in absolute terms (Panel~B).  This pattern---fewer cyclone-strength observations overall, driven overwhelmingly by fewer \C1 observations---suggests not only promotions from weaker to stronger categories, but also fewer storms reaching or maintaining minimal \C1 cyclone strength to begin with.

With the extended data, the picture changes.  Total \C{1+} observations per year are nearly flat ($-1$\%), and the per-year rates of \C2 through \C4 observations all \emph{increase}.  The \C1 decline persists but is smaller ($-50$/yr vs.\ $-59$/yr).  The extended-period changes are thus more consistent with a genuine shift from weaker to stronger cyclones.

Panel~C reports Kossin's $R$ statistic and its decomposition into category shares.  The \C5 share declines in both periods, so the intensification is not monotonic across categories; it is concentrated in \C3 and~\C4.  Using HURSAT v7 data, $\Delta R$ is 0.026\footnote{\textcite{kossin2020global} reports $P_\text{maj}$ (= $R$) rising from approximately 0.355 to 0.388 in the full-resolution v6 data, giving $\Delta R \approx 0.033$ (found in Table~1 and Figure~1 of \textcite{kossin2020global}).  The v6 supplemental data (subsampled to 6-hourly) give $P_\text{maj} = 0.338 \to 0.372$, i.e., $\Delta R = 0.034$.  The v7 value of 0.026 is smaller, reflecting the ADT version shift rather than a methodological discrepancy.} for the Kossin period and rises to 0.046 in the extended sample.

\begin{figure}[ht]

  \figcaption{fig:deltacyc}{Changes in Tropical Cyclone Intensities Per Year, Late Minus Early}

  \begin{ctabular}{c}
    \textbf{Kossin Years (1979--2017)} \\
    \addlinespace
    \includegraphics[width=0.82\textwidth]{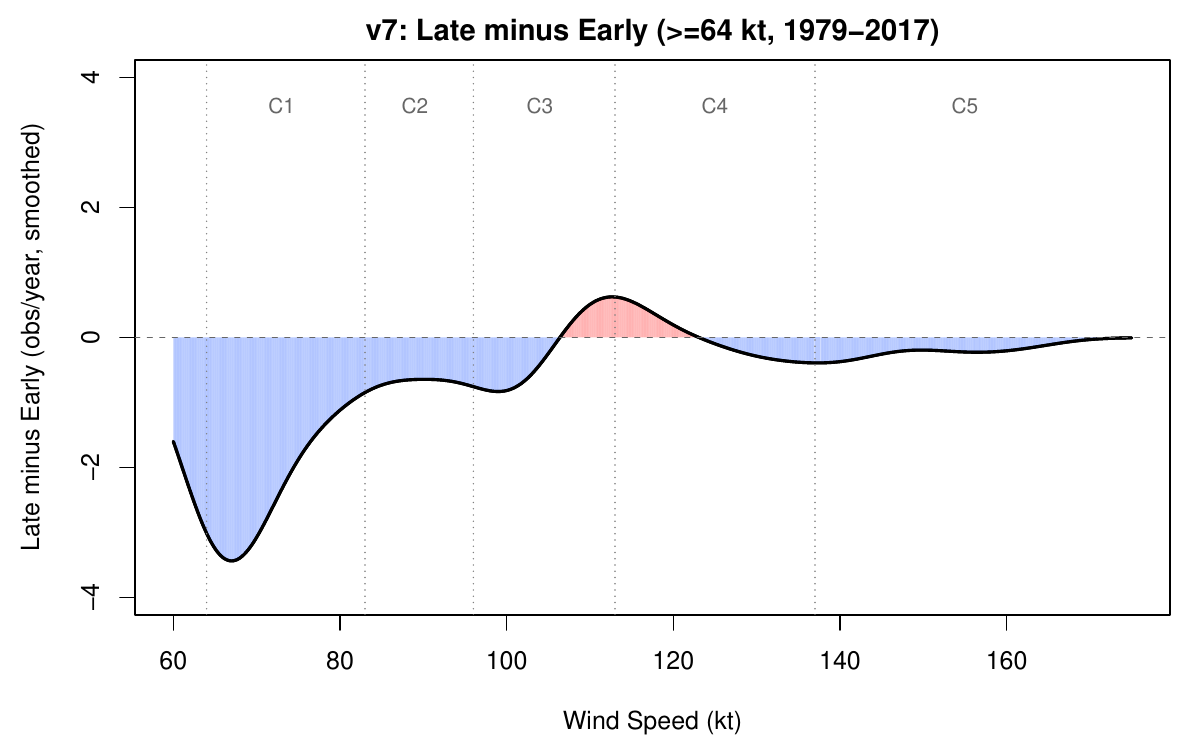} \\
    \addlinespace
    \textbf{Extended Years (1979--2023)} \\
    \includegraphics[width=0.82\textwidth]{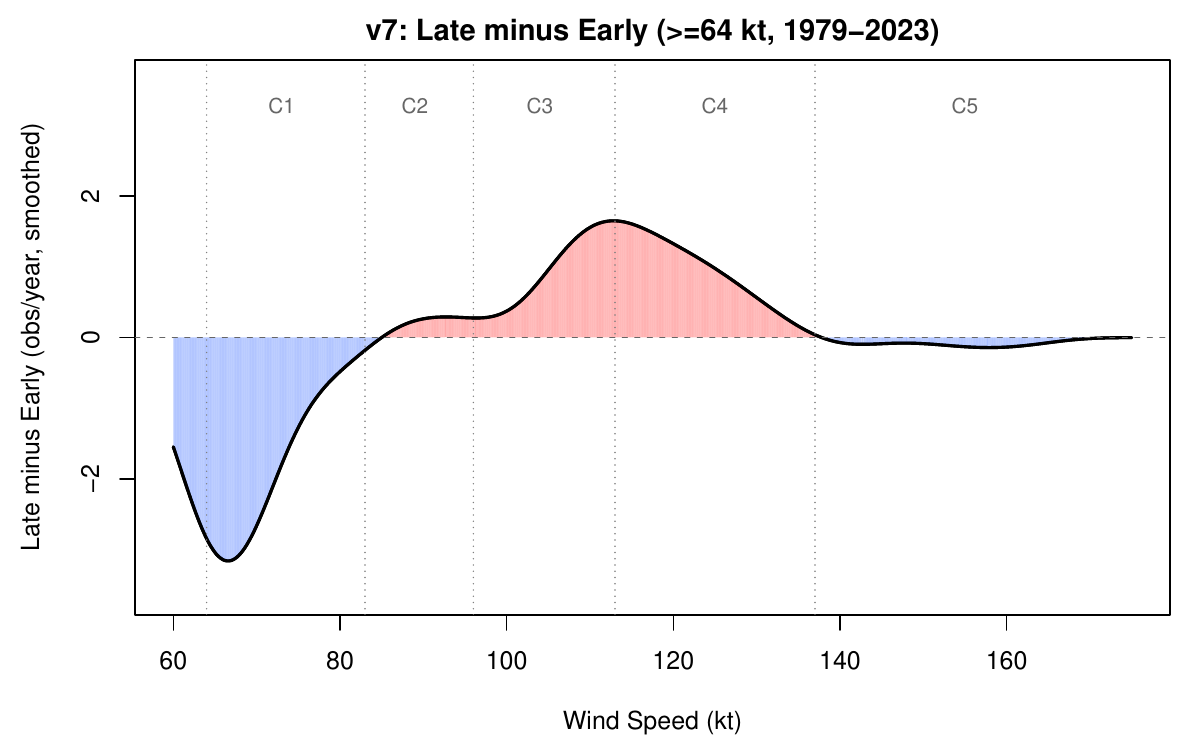}
  \end{ctabular}

  \small

  \explain{Top: Kossin's period (1979--2017, split 1997/98).  Bottom: Extended (1979--2023, split 2000/01).  Blue (red) shading: the late period has \emph{fewer} (\emph{more}) cyclone observations than the earlier period.  Unlike Table~\ref{tbl:counts}, the figure uses the full 1-kt wind-speed resolution, kernel-smoothed with a 5-kt Gaussian bandwidth, rather than binning into Saffir-Simpson categories.}

\end{figure}

\insfig{fig:deltacyc}

Figure~\ref{fig:deltacyc} makes this point visually obvious.  These are kernel density estimates of the wind-speed distribution, smoothed at 5-kt bandwidth, computed separately for each half-sample and differenced.  The reported units are \emph{changes in late vs early cyclones, in observations per year}.  For 1979--2017 (top panel), the blue deficit at \C1 wind speeds ($\sim$65--80~kt) is large---peaking at 3.4~fewer observations per year---while the red surplus in the \C3--\C4 range is comparatively small.  Almost all of the relative intensification comes from fewer \C1 observations.  In raw terms, the late half simply has fewer cyclone observations at nearly every intensity below about 105~kt.

For 1979--2023 (bottom panel), the picture changes materially: the \C3--\C4 surplus grows to roughly match the \C1 deficit.  The recent years (2018--2023) contribute genuinely more major cyclone observations, not just fewer weak ones.

\subsection{Trend Analysis}

\instbl{tbl:trends}

\begin{table}[ht]

  \tblcaption{tbl:trends}{Annual category share trends: Theil-Sen slopes and Mann-Kendall $P$-values}

  \begin{ctabular}{cr Rn Rn Rn}
    \toprule
              &          &  \mc6c{Slope Per Decade}                                                                 \\
    \cmidrule(lr){3-8}
              &          & \mc2c{Kossin v6} & \mc4c{HURSAT v7}                                                      \\
    \cmidrule(lr){3-4} \cmidrule(lr){5-8}
              &          & \mc2c{1979--2017}    & \mc2c{1979--2017} & \mc2c{1979--2023}                                         \\
    \midrule
    (\C1)     & $\nk{1}$ & -0.024\tstar     & 0.002\;***    & -0.022\tstar & 0.007\;*** & -0.025\tstar & 0.001\;*** \\
    (\C2)     & $\nk{2}$ & +0.005           & 0.236         & +0.002       & 0.411      & +0.005\ostar & 0.077\;*   \\
    \addlinespace
    (\C3)     & $\nk{3}$ & +0.014\tstar     & 0.004\;***    & +0.011\dstar & 0.031\;**  & +0.010\dstar & 0.014\;**  \\
    (\C4)     & $\nk{4}$ & +0.011\dstar     & 0.045\;**     & +0.011\ostar & 0.066\;*   & +0.012\dstar & 0.011\;**  \\
    (\C5)     & $\nk{5}$ & -0.002           & 0.411         & -0.002       & 0.483      & -0.003       & 0.166      \\
    \addlinespace \midrule \addlinespace
    (C3+/C1+) & $R$      & +0.023\tstar     & 0.005\;***    & +0.020\dstar & 0.042\;**  & +0.020\dstar & 0.010\;**  \\
    \bottomrule
  \end{ctabular} {\vspace{-1em}\scriptsize \ostar\quad: $p<0.10$.  \dstar\quad: $p<0.05$.  \tstar\quad: $p<0.01$.}

  \explain{Theil-Sen slopes (per decade) of annual category shares $\N{k}/\N{1+}$. Significance is Mann-Kendall.  The slope of $R$ approximately equals the sum of the \C3, \C4, and \C5 slopes. (The equality is exact for OLS but only approximate for Theil-Sen, since the median of a sum need not equal the sum of medians.)}
\end{table}

\textcite{kossin2020global} also analyze the time trend in their $R$ statistic, rather than merely the first vs. second half difference.  To do so, they use 3-year smoothed time series and Theil-Sen/Mann-Kendall nonparametric tests.\footnote{Materials and Methods explains that this is not necessary.  Inference from plain OLS regressions is virtually identical.  Using annual data rather than triads also allows using the 2024 observation.  The results are unchanged.} Table~\ref{tbl:trends} applies the tests to each category share separately.  Again, it shows which categories drove their overall $R$ trend and at what significance level.

The \C1 share declines at 2.5 percentage points per decade, the most significant individual trend ($P = 0.001$).  The \C3 and~\C4 shares each increase at about 1~percentage point per decade, both significant at the 5\% level.  The \C5 share shows an insignificant decline.

The middle column of Table~\ref{tbl:trends} isolates the effect of the data version change from the effect of extending the record.  Over the same 1979--2017 period, v6 and v7 produce similar slopes (e.g., $R$: $+0.023$ vs.\ $+0.020$/decade), confirming that the ADT version shift does not alter the trend.  Significance is somewhat lower in v7 for this period ($P = 0.042$ vs.\ $P = 0.005$), possibly because v7 produces more observations near the classification boundary, adding noise to the shares.

The extended record strengthens the overall statistical significance further.  Using Kossin's triad smoothing, \textcite{kossin2020global} report Mann-Kendall $P = 0.02$ for the $R$ trend over 1979--2017; extending through 2023 with the same method yields $P = 0.003$.  Using annual data instead of triads (Table~\ref{tbl:trends}), $P = 0.010$.  The two most recent 3-year periods (centered on 2019 and 2022) produce $R$ values of 0.496 and 0.489---among the highest in the record---reinforcing the trend rather than diluting it.

In sum, the data have now shifted from ``just fewer weak storms'' to ``fewer weak storms \emph{and} more strong ones.''

\subsection{Basin Decomposition}

\instbl{tbl:basins}

\begin{table}[ht]
  \tblcaption{tbl:basins}{Per-Basin $R$ half-sample differences and trends (v7, 1979--2023)}

  \begin{ctabular}{ll r rrR Rn}
    \toprule
    & & & \mc3c{Late Minus Early} & \mc2c{Theil-Sen} \\
    \cmidrule(lr){4-6} \cmidrule(lr){7-8}
                       & Basin & Obs/yr & $R_\text{early}$ & $R_\text{late}$ & $\Delta R$ & Slope/dec & MK $P$     \\
    \midrule
    North Atlantic     & NA    & 117    & 0.347            & 0.474           & +0.126     & +0.063\tstar & 0.002\;*** \\
    Eastern Pacific    & EP    & 226    & 0.381            & 0.404           & +0.023     & +0.010       & 0.353      \\
    Western Pacific    & WP    & 279    & 0.508            & 0.515           & +0.007     & +0.004       & 0.807      \\
    Northern Indian    & NI    &  15    & 0.292            & 0.412           & +0.121     & +0.000       & 0.549      \\
    Southern Indian    & SI    & 101    & 0.431            & 0.506           & +0.076     & +0.056\tstar & 0.001\;*** \\
    South Pacific      & SP    &  74    & 0.405            & 0.467           & +0.062     & +0.026       & 0.237      \\
    \midrule
    All                &       & 812    & 0.427            & 0.473           & +0.046     & +0.020\dstar & 0.010\;**  \\
    \bottomrule
  \end{ctabular}
  {\vspace{-1em}\scriptsize \ostar\quad: $p<0.10$.  \dstar\quad: $p<0.05$.  \tstar\quad: $p<0.01$.}

  \explain{Early: 1979--2000, Late: 2001--2023.  Obs/yr is the mean annual \C{1+} observation count over 1979--2023.  $R = \N{3+}/\N{1+}$.  Theil-Sen slope per decade with Mann-Kendall $P$-value.  NI averages only 15 \C{1+} observations per year.  Its large $\Delta R$ reflects small-sample volatility rather than a robust trend, as the near-zero Theil-Sen slope confirms.}
\end{table}

Table~\ref{tbl:basins} decomposes the global result by basin.  With fewer observations per basin, statistical significance is harder to achieve.  Nevertheless, the trend is significant in the North Atlantic ($P = 0.002$) and Southern Indian ($P = 0.001$) basins.  Both improve markedly relative to the shorter record: using v7 triads over 1979--2017, they were only marginally significant ($P = 0.059$ and $P = 0.077$, respectively).  The Western Pacific---the largest basin by observation count---shows essentially no trend.

Not tabulated, the mechanism behind the rising $R$ differs across basins.  In the North Atlantic, both weak and strong cyclone observations increase, but \C{3+} observations grow far faster, producing the largest $\Delta R$ of any basin (+0.126).  In the Southern Indian, weak cyclone observations are roughly flat while \C{3+} observations rise---a pattern closer to pure intensification.  In the Eastern and South Pacific, all categories decline, but weak observations decline faster, raising the ratio despite fewer storms overall.  In the Western Pacific, all categories are essentially trendless.

\section{Discussion}

The declining \C1 share even in the newer data raises a natural question: is it real weather or a measurement artifact?  ADT-HURSAT was designed for homogeneity, but \C1 observations sit near the algorithm's classification boundary (64~kt), where small changes in satellite resolution or calibration could have disproportionate effects.  Whether the decline reflects fewer tropical storms reaching cyclone strength, faster intensification through \C1, or subtle changes in satellite coverage remains an open question that this decomposition sharpens but cannot resolve.

Geographically, the trend is concentrated in the North Atlantic ($P = 0.002$) and Southern Indian Ocean ($P = 0.001$).  The Western Pacific, despite contributing the most observations, shows essentially no trend.  Whether this reflects genuine regional differences in environmental forcing or simply statistical power limitations is unclear.  No multiple-testing correction is applied to per-category or per-basin $P$-values; the decomposition is exploratory, and the global trend significance ($P = 0.010$) does not depend on it.

In sum, this brief report has shown that the Kossin major-cyclone ratio $R = \N{3+}/\N{1+}$ decomposes exactly into five additive category shares.  For 1979--2017, the trend was primarily driven by a declining share of \C1 observations, not by increasing shares of intense storms.  For the extended record through 2023, genuine intensification in \C3 and \C4 has emerged, roughly matching the weak-cyclone decline in magnitude.  The decomposition provides a transparent accounting of what drives the ratio trend and clarifies what physical mechanisms need to be invoked.


\section{References}
\printbibliography[heading=none]

\clearpage
\appendix

\section{APPENDIX: Materials and Methods}

\subsection{Data}

\paragraph{ADT-HURSAT v6 (Kossin's original).}  \textcite{kossin2020global} applied an older version of the Advanced Dvorak Technique (ADT) to HURSAT~v6 satellite imagery.  Wind speeds were recorded in 5-kt bins (25, 30, \ldots, 170~kt), and cyclone thresholds followed the rounded Saffir-Simpson scale: \C1~$\ge$~65, \C2~$\ge$~85, \C3~$\ge$~100, \C4~$\ge$~115, \C5~$\ge$~140~kt.  The dataset covered 4,180 storms from 1978--2017.  The original full-resolution ($\sim$3-hourly) data were never publicly released.  The PNAS supplemental files (Datasets~S1--S9) contained the same storms subsampled to 6-hourly synoptic intervals (approximately 134,000 total observations across all wind speeds).

\paragraph{ADT-HURSAT v7b (updated).}  The updated dataset applies ADT version~9.0 to HURSAT~v07b imagery.\footnote{NCEI Accession~0307249.  Individual storm files are hosted on Google Cloud Storage at \texttt{gs://noaa-ncei-ipg/datasets/hursat/adt/history-files/}.}  Wind speeds are continuous (1-kt resolution), with standard Saffir-Simpson thresholds: \C1~$\ge$~64, \C2~$\ge$~83, \C3~$\ge$~96, \C4~$\ge$~113, \C5~$\ge$~137~kt.  The dataset covers 4,852 storms from 1978--2025, with observations at approximately 3-hourly resolution.

\paragraph{Key differences.}  ADT~v9.0 yields systematically higher wind speeds than the older algorithm: 68\%~more observations exceed the cyclone threshold (64~kt) in v7 than in v6.  This level shift is large, but the relative variation is preserved: Pearson correlations between v6 and v7 are 0.965 for \C{1+} observation counts by basin$\times$year ($N = 234$), 0.975 for \C{3+} counts, and 0.902 for the \C{3+}/\C{1+} ratio.  At the 3-year aggregation level used by Kossin, these correlations rise to 0.985, 0.993, and 0.929 respectively.  Trend directions and relative magnitudes are robust to the version change.

\needspace{4\baselineskip}
\subsection{Statistical Methods}

The statistical methods in \textcite{kossin2020global} are valid.  Nevertheless, for the share decomposition, simpler methods suffice because the data are well-behaved.

Kossin tests early-vs-late differences via non-overlapping confidence intervals rather than a direct test of $\Delta R$.  This is conservative: non-overlapping CIs are sufficient but not necessary for significance at the stated level.  A direct two-proportion $z$-test (with the same effective-$N$ adjustment) would yield equal or smaller $P$-values.

\begin{description}

\item[Trend regressions with outlier robustness (Theil-Sen).]  \textcite{kossin2020global} uses the Theil-Sen estimator (median of all pairwise slopes), which has a breakdown point of $\sim$29\%---it tolerates up to 4 arbitrary outliers among 15 triad points.  However, the annual ratio data contains no contaminated outliers.  All extreme values seem ``normal'' (e.g., real weather in North Indian Ocean in 2019, and South Pacific Ocean in 1983).  Accordingly, the Theil-Sen and OLS slopes are nearly identical: $+0.020$ vs.\ $+0.019$/decade for the annual ratio over 1979--2023.  We retain Theil-Sen for consistency with Kossin but note that the choice is immaterial.  An advantage of not using triads is that the final year of data becomes usable, too.  The results are unchanged.

\item[Serial correlation.]  \textcite{kossin2020global} adjusts confidence intervals for autocorrelation along tropical cyclone tracks (decorrelation time $\sim$18~hours, reducing effective sample size by $\sim$3$\times$).  This adjustment applies to the early-vs-late CI comparison but is irrelevant for the annual share trend tests.  The Durbin-Watson statistic on annual ratio residuals is 2.12 (ideal~$=$~2.0), and the lag-1 autocorrelation of OLS residuals is $-0.06$.  Annual observations are essentially independent, so no degrees-of-freedom correction is needed.  Kossin's own Durbin-Watson test on 3-year triad points also found no significant autocorrelation.

\item[Mann-Kendall vs. OLS.]  The Mann-Kendall test (nonparametric monotonic trend test) pairs naturally with Theil-Sen.  For the annual ratio, Mann-Kendall gives $P = 0.010$ and OLS gives $P = 0.003$---both clearly significant.  Per-category results are also consistent across methods (Table~\ref{tbl:trends}).

\end{description}

\needspace{4\baselineskip}
\subsection{Theil-Sen Slopes, Different Categories}

\insfig{fig:ts}

Figure~\ref{fig:ts} shows the annual time series for each category share and for the composite ratio.  The central test in \textcite{kossin2020global} was for the bottom right diagram, indicating an increase in relative cyclone intensities of approximately 2 percentage points per decade.

\begin{figure}[p]
  \figcaption{fig:ts}{Annual category shares and Kossin ratio, 1979--2023, with Theil-Sen trend lines}

  \includegraphics[width=\textwidth]{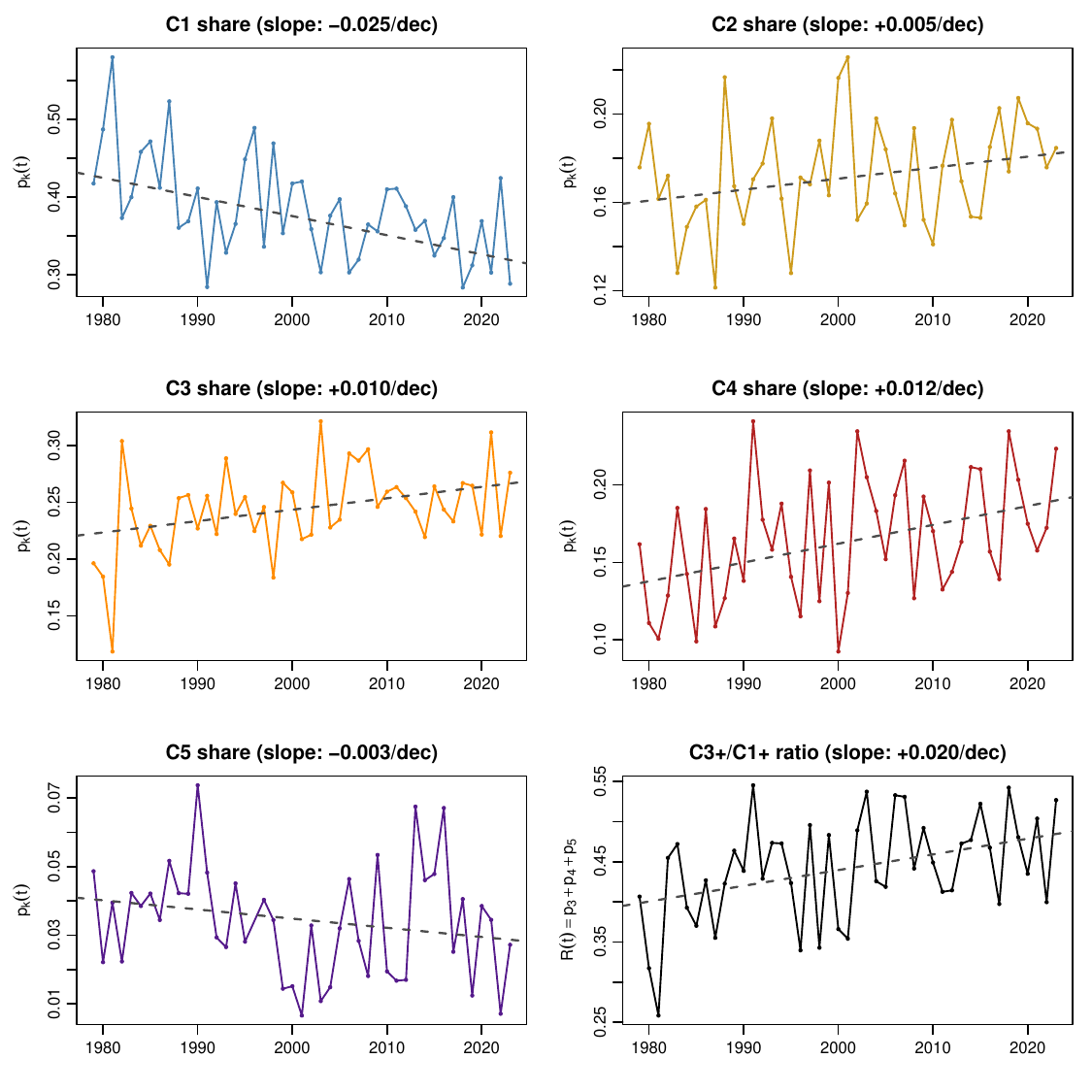}

  \explain{Each panel shows the annual share $\N{k}(t)/\N{1+}(t)$ for one Saffir-Simpson category. The bottom right panel shows the composite ratio $R(t) = \nk{3} + \nk{4} + \nk{5}$.  Theil-Sen trend lines are overlaid.}
\end{figure}

\end{document}